\documentstyle[12pt,aasms4]{article}
\lefthead{Kulkarni, Fall, \& Truran}
\righthead{Abundances in Damped Lyman-alpha Galaxies}

\begin{document}

\title{ABUNDANCE PATTERNS OF HEAVY ELEMENTS \\
IN DAMPED LYMAN-ALPHA GALAXIES}

\author{Varsha P. Kulkarni \altaffilmark{1,2} and S. Michael Fall}

\affil{Space Telescope Science Institute, 3700 San Martin 
Drive, Baltimore, MD 21218, U.S.A.} 

\and

\author{J. W. Truran \altaffilmark{3} }
\affil{Department of Astronomy \& Astrophysics, 
University of Chicago, \\
5640 S. Ellis Ave.,Chicago, IL 60637, U.S.A.}

\altaffiltext{1} {Also, Department of Astronomy \& Astrophysics, 
University of Chicago, 5640 S. Ellis Ave.,Chicago, IL 60637, U.S.A.} 
\altaffiltext{2} {Current Address: Steward Observatory, University of 
Arizona, Tucson, AZ 85721, U.S.A.}
\altaffiltext{3} {Also, the Enrico Fermi Institute}

\begin{abstract}
We present a quantitative analysis of the abundances of heavy elements 
in damped Ly$\alpha$ galaxies in the sample of Lu et al. (1996). 
In particular, we compare the observed gas-phase abundances with those
expected when the intrinsic (i.e., nucleosynthetic) pattern is the same 
as that in either the Sun or in Galactic halo stars and when the 
depletion pattern is the same as that in the warm Galactic interstellar 
medium, but with various values of the dust-to-metals ratio.
We find that the observations are equally consistent with the solar and 
halo-star intrinsic patterns and that they favor some depletion, the 
typical dust-to-metals ratio being $40\%-90\%$ of 
that in the Milky Way today.
However, neither intrinsic pattern matches the observations perfectly. 
For the solar pattern, the discrepancy is mainly with [Mn/Fe], while for 
the halo-star pattern, the discrepancy is with [Zn/Fe], [Ni/Fe], and 
possibly [Al/Fe]. Our analysis does not support the claim by Lu 
et al. that the damped 
Ly$\alpha$ galaxies have halo-star abundance patterns and no 
dust depletion.
\end{abstract}


\section{INTRODUCTION}

The abundances of heavy elements in damped Ly$\alpha$ (DLA) galaxies 
provide important clues about the histories of metal production and 
star formation in these objects (Lanzetta, Wolfe, \& Turnshek 1995; 
Pei \& Fall 1995; Timmes, Lauroesch, \& Truran 1995).
Many of the observational studies to date have focussed on Zn and Cr 
(Pettini et al. 1997, and references therein). These elements track Fe 
almost perfectly in Galactic stars with $-3\lesssim$ [Fe/H] $\lesssim 0$ 
(see, e.g., Wheeler, Sneden, \& Truran 1989). Since Zn 
is relatively undepleted while Cr is strongly depleted in the Galactic  
interstellar medium (ISM) (e.g., Savage \& Sembach 1996), [Zn/H] should 
be a reliable indicator of the 
metallicity and [Cr/Zn] should be an indicator of the dust content.
Measurements of Zn and Cr show that the typical metallicity and 
dust-to-gas ratio in DLA galaxies at $\bar z \approx 2$ are 
$\sim 1/10$ the corresponding values in the local ISM (Pettini et al.
1994). The low dust-to-gas ratio is also consistent with the mild 
preferential reddening of quasars with DLA galaxies in the foreground 
(Pei, Fall, \& Bechtold 1991). These results suggest that the typical 
dust-to-metals ratio has remained roughly constant (to within factors 
of 2 or so) since $z\approx2$, and that the dust is produced in step 
with the metals.

Lu et al. (1996) recently obtained high-resolution,
high-S/N spectra of 14 DLA galaxies using the Keck Telescope. 
Combining these and similar observations of nine DLA galaxies 
from the literature, they concluded that the DLA galaxies have the 
nucleosynthesis
pattern expected purely from Type II supernovae, similar to that found
in Galactic halo stars, and show no evidence of dust depletion. However, 
as we show 
in this Letter, the Lu et al. interpretation is not unique. The main 
reason for this is that the differences in the relative 
abundances caused by Type I and Type II supernovae are comparable to 
the differences caused by even small amounts of dust depletion. We
demonstrate this explicitly by comparing the observed relative 
abundances of different pairs of elements with the values expected
for different assumptions about the nucleosynthesis pattern and
the dust-to-metals ratio. Our conclusions are similar to those reached 
by Lauroesch et al. (1996), who considered an earlier set of data on 
many different elements, and Pettini et al. (1997), who considered a 
homogeneous set of data on Zn and Cr. 

\section{METHOD AND RESULTS}

To derive the expected relative gas-phase abundances in 
terms of the dust-to-metals ratio, we start with the basic equation 
that relates the abundances of atoms of any element X in the gas and 
solid phases ($X^{gas}$ and $X^{sol}$) to the total abundance 
($X^{tot} = X^{gas} + X^{sol}$). 
We assume that the relative total abundance of any 
two elements X and Y 
is the same as that in a known reference pattern, which may be chosen 
later to be either the solar pattern or the halo-star pattern, i.e.,
\begin{equation}
{X^{tot} / {Y^{tot}}} = ({X^{tot} / {Y^{tot}}})_{ref}. 
\end{equation}
Next, we assume that the DLA galaxies have 
the same relative dust-depletion pattern, i.e., the same composition 
of dust grains, as a reference ISM and differ only in the absolute 
number of dust grains per unit mass of the ISM. This gives  
\begin{equation}
{X^{sol} / {Y^{sol}}} = ({X^{sol} / {Y^{sol}}})_{ref},
\end{equation}
and hence, 
\begin{equation}
{X^{gas} \over {Y^{gas}}} = {{( X^{tot} / Y^{tot} )_{ref} - 
(Y_{sol}/Y_{tot}) (X^{sol}/Y^{sol})_{ref} } \over {1-
(Y_{sol}/Y_{tot})}}.
\end{equation}

For simplicity, we take the reference depletion pattern to be that 
in the Galactic ISM and assume that the total  
abundance of each element in the Galactic ISM equals the solar 
abundance. {\footnote {Some recent observations indicate nonsolar 
abundances in the Galactic ISM. However, since these results are not 
yet definitive and 
may be caused by purely local effects (Savage \& Sembach 1996), we 
continue to adopt the solar values.}}   
Denoting the gas-phase abundances, relative to the solar abundances, 
of X and Y in the Galactic ISM as 
$\Delta_{X\,G} = (X^{gas}/H)_{G}/(X/H)_{\odot}$ and 
$\Delta_{Y\,G} = (Y^{gas}/H)_{G}/(Y/H)_{\odot}$, we can now reexpress 
equation (3) in the form
\begin{equation}
 {X^{gas} \over {Y^{gas}} }  =  { {( X^{tot} / Y^{tot} )_{ref} 
- (R/R_{G})_{Y} (X/Y)_{\odot} (1-\Delta_{X\,G})} \over 
{[1- (R/R_{G})_{Y}(1 - \Delta_{Y\,G})] }} ,
\end{equation}
where 
\begin{equation}
(R/R_{G})_{Y} \equiv 
(Y^{sol}/Y^{tot})/(Y^{sol}/Y^{tot})_{G} = (Y^{sol}/Y^{tot})/ 
(1 - \Delta_{Y\,G})
\end{equation} 
denotes the dust-to-metals ratio of Y   
relative to that in the Milky Way. We use the notation [X/Y] to represent 
$\log \{ {(X^{gas}/Y^{gas}) / {(X/Y)_{\odot}}} \}$ in DLA galaxies. 
In the following, we take the reference element Y to be Fe or Zn, since 
these have both been used as metallicity indicators for DLA galaxies 
(although Fe has the serious disadvantage that it depletes easily onto 
dust grains). The dust-to-metals ratio for any other element Y is 
related to the dust-to-metals ratio for Fe by the relation 
$(R/R_{G})_{Y}  = (R/R_{G})_{Fe} {(Y/Fe)_{\odot} / 
{(Y^{tot}/Fe^{tot})_{ref}}}$. 
Since the halo-star abundances of Zn, Cr, and Ni relative to Fe are very 
close to the solar values for $-3 \lesssim$ [Fe/H] $\lesssim -1$, we 
have $(R/R_{G})_{Zn} = 
(R/R_{G})_{Fe}$ for both the halo-star and the solar intrinsic abundance 
patterns (and similarly for Cr or Ni). We therefore 
use the symbol $R/R_{G}$ to 
denote $(R/R_{G})_{Zn}$ and $(R/R_{G})_{Fe}$. 
Equation (4) shows that 
the gas-phase abundances with respect to Fe or Zn are determined 
by $R/R_{G}$ alone. 

In this Letter, we focus on the abundances of S, Si, Al, Zn, Fe, Cr, 
Ni, and Mn. {\footnote{ We do not use N and O because very few 
reliable measurements exist for these elements due to problems of 
line saturation and blending with the Ly$\alpha$ forest. The  
existing measurements also show a large scatter in [N/O] (Green et al. 
1995; Lu et al. 1996; Molaro et al. 1996). }} For each of these 
elements, we choose the input quantities in equation (4) as follows. 
We adopt the solar abundances from Anders \& Grevesse (1989). 
For the relative halo-star abundances [X/Zn]$_{halo}$ or
[X/Fe]$_{halo}$, we use the following  
values: +0.4 dex for Si and S; -0.3 dex for Al; 
-0.25 dex for Mn; 0.0 dex for Cr, Ni, Fe, and Zn. 
These are representative 
values based on observations of Galactic halo stars with 
$-3 \lesssim $ [Fe/H] $ \lesssim -1$ (see, e.g., 
Gratton \& Sneden 1987, 1988, 1991; Magain 1989; 
Ryan, Norris, \& Bessell 1991; Sneden, Gratton, \& Crocker 1991; 
McWilliam et al. 1995; Ryan, Norris, \& Beers 1996).  
Our assumed value for Al is perhaps the least certain, since the 
abundance of this element shows considerable scatter in halo stars. 
For the Galactic depletions 
$\Delta_{X G}$, we adopt the values observed in the warm, diffuse ISM, 
since these seem appropriate for comparison with the DLA galaxies, 
which in most cases show almost no detectable H$_{2}$ 
[Levshakov et al. 1992; but see  
Ge \& Bechtold 1997 for a case with relatively high $f($H$_{2})$]. 
For S, Mn, Cr, Si, Fe and Ni, we adopt the warm disk Galactic 
depletions from Table 6 of Savage \& Sembach (1996). For Zn, we use 
$\log \Delta_{X \,G} =  -0.20$ (Roth \& Blades 1995; Sembach et al. 
1995), and for Al, we use  $\log \Delta_{X \,G} =  -1.16$  
(Barker et al. 1984, corrected for the revised oscillator 
strength from Morton 1991). The value for Al is somewhat uncertain 
because of line saturation effects.
 
We now compare the predictions from equation (4) with the data 
for the 23 DLA galaxies from Table 16 of Lu et al. (1996). These 
absorbers have $0.7 \le z \le 4.4$ and 
$20.0 \le \log N_{\rm {HI}} \le 21.7$. 
Unfortunately, simultaneous measurements of more than three elements 
exist for only a third of these objects. Hence, we exploit the relatively 
large sample to infer the typical abundance pattern in DLA galaxies 
rather than the pattern in any particular system. 
In Figures 1 and 2, we plot the 
logarithmic gas-phase abundances of S, Mn, Cr, Si, Fe (or Zn), Ni, 
and Al, relative to Zn or Fe, in increasing order of condensation 
temperature. Following Lu et al., we assume 
that the absorbers are mostly neutral. In both Figures, the 
upper panels are scaled to the solar abundances 
[i.e., $(X^{tot}/Y^{tot})_{ref} = (X/Y)_{\odot}$], while the lower 
panels are scaled to the halo-star abundances 
[i.e., $(X^{tot}/Y^{tot})_{ref} = (X^{tot}/Y^{tot})_{halo}$]. 
The light vertical segments show the full range of the actual  
measurements for the DLA galaxies. The heavy filled circles and 
vertical bars show the medians of the  
observations (including upper and lower limits) and the standard 
errors in the medians (see, e.g., Sachs 1984). 
These errors may 
be underestimates in some cases, since they include only the 
statistical, not the measurement, uncertainties (which are typically 
$\pm 0.15$ dex or less). 
We consider the medians rather than the means because the sample is 
small, and because they allow the use of limits as well 
as actual measurements. This procedure works in most cases  
because the element ratios are available for at least 
three absorbers and typically many more. However, the ratios  
[S/Zn], [S/Fe], [Al/Zn], [Al/Fe], and [Mn/Zn] are available for 
fewer than three absorbers, not enough to compute reliable median 
values. Therefore, for these, we show the individual 
measurements with open circles and the upper or lower 
limits with open triangles. 

The lines in Figures 1 and 2 represent the expected gas-phase 
abundances for different assumed values of the dust-to-metals 
ratio: $R/R_{G} = 0$ (light horizontal lines), 
0.3 (short-dashed lines), 0.6 (long-dashed 
lines), and 0.9 (dot-dashed lines). In both Figures, we have assumed 
a solar intrinsic pattern in the top panels and a halo-star-like 
intrinsic pattern in the bottom panels. If DLA galaxies possessed no 
dust, as argued by Lu et al. (1996), 
then the observed relative abundances for all elements would lie 
close to the light horizontal line in each Figure. Clearly, this 
is not the case. Based on a Kolmogorov-Smirnov test, the observed 
deviation in Figure 1 from the case of zero dust is significant at the 
level of $> 99 \%$ with or without the inclusion of S and Mn. 
For Figure 2, the significance of the deviation is $\approx 99 \%$ 
including S and Al and $\approx 97 \%$ excluding them. Thus, 
we find evidence for dust depletion irrespective of whether the 
intrinsic abundances are assumed solar or halo-star-like. The solar 
pattern with 
$R/R_{G} \approx 0.4-0.9$ is roughly consistent with the observations 
of most of the 
elements. This range of $R/R_{G}$ agrees nicely with 
the result of Pettini et al. (1997) based on Zn and Cr 
alone. {\footnote{ However, our estimates of the typical metallicity 
and dust-to-gas ratio in the DLA galaxies from the Lu et al. (1996) 
sample are $\approx 1/10 Z_{\odot}$ and $\approx 1/15$ of the Galactic 
value (Kulkarni 1996), i.e., 
$\approx 1.5-2$ times higher than the estimates by Pettini et al. (1997). 
This slight difference arises because we have included the weak 
depletion of Zn ($\approx 0.2$ dex) in the warm Galactic ISM, 
while Pettini et al. have not. }}

Closer inspection shows that a single value of $R/R_{G}$ cannot 
simultaneously fit the observations of all the elements. For example, 
in Figure 1{\it b}, S and Si cannot be fitted at all. In 
Figure 2{\it a}, Mn is the only discrepant element. There is no 
discrepancy for Ni in 
the sense that there is for Mn, because the fit for the former  
improves as the dust-to-metals ratio increases. The discrepancy for 
[Mn/Fe] is the main point used 
by Lu et al. (1996) to argue that DLA galaxies are dust-free. 
However, since Mn and Fe are both strongly 
depleted even in the warm Galactic ISM, they are less reliable 
indicators of the dust content than are Zn and Mn, or Zn and Fe, or 
Zn and Cr. In fact, from Figure 1, the data for [Mn/Zn] and [Fe/Zn] 
both require some dust depletion. 
In Figure 2{\it b}, Zn and Ni require more depletion 
compared to S, Mn, and Si, while Al cannot be fitted at all.  
This is difficult to reconcile with a purely halo-star 
abundance pattern with no dust. We have also 
performed calculations with a depletion pattern similar to that in 
cold clouds in the Galactic ISM 
and find that it also fails to give simultaneous agreement for all 
elements, for solar or halo-star abundance patterns. 
This suggests that the dust depletion pattern in DLA galaxies 
may differ from that in the Galactic ISM . In any case, 
we conclude that dust depletion can easily 
mask or confuse the underlying nucleosynthetic pattern, in agreement 
with the suggestion by Lauroesch et al. (1996). 

\section{DISCUSSION} 

The evidence presented here that the DLA galaxies at $z \approx 2-4$ 
contain small but significant amounts of dust is consistent with 
several other observations. 
These include the mild reddening of quasars with DLA galaxies in the 
foreground (Pei et al. 1991) and the mild depletion of Cr relative to 
Zn in DLA galaxies (Pettini et al. 1994, 1997). 
Another relevant observation is that, in most cases, the DLA galaxies 
emit weakly if at all in the Ly$\alpha$ line (Lowenthal et al. 1995 and 
references therein). This is also true of the starburst galaxies found 
at $z \approx 3$ by Steidel et al. (1996), which would almost certainly be 
classified as DLA galaxies, if they could be observed in absorption. 
The simplest explanation for these observations 
is that Ly$\alpha$ photons 
are produced within the galaxies (in the recombinations following
H ionization by young stars and other sources) but are subsequently
absorbed by dust grains before escaping. There are other possibilities, 
but they involve 
either special viewing angles or times and would not apply to a 
population of randomly oriented galaxies of various ages (Charlot \& 
Fall 1993, and references therein).

The presence of some dust in DLA galaxies is certainly plausible. Dust 
is produced in the cool envelopes of
intermediate-mass stars and in the dense shells of supernova 
remnants, although it is not clear how long it survives once it enters 
the ISM (which depends on the unknown intensity of the 
ultraviolet radiation field, frequency of shocks, and so forth). 
Similarly, it seems reasonable that the DLA galaxies have neither 
purely solar nor purely halo-star abundance patterns. 
The DLA galaxies at $z\approx2$ are 1--2 Gyr older than those at 
$z\approx4$, a range in ages probably sufficient to include enrichment 
by both Type I and Type II supernovae. The large scatter in the 
metallicities of the DLA galaxies at each redshift (a factor 
of $\approx 15-30$ as measured by [Zn/H] or [Fe/H]) also indicates that 
they are observed in very different stages of chemical evolution and 
therefore that they could be expected to exhibit a mixture of both 
solar and halo-star abundance patterns. We note, however, that DLA 
galaxies with the highest metal and hence dust content tend to be underrepresented in samples derived from optically 
selected quasars (Fall \& Pei 1993).

Lu et al. (1996) reached different conclusions because they disregarded 
Zn as a metallicity indicator. 
As we have already emphasized, the rationale for choosing Zn is that 
it tracks Fe almost perfectly in Galactic stars over a wide range of 
metallicities and is nearly undepleted in the Galactic ISM. 
In contrast, Fe, the metallicity indicator preferred by Lu et al., is 
highly depleted in the Galactic ISM. As justification for ignoring Zn, 
Lu et al. appealed to theoretical models of Type II supernovae by 
Woosley \& Weaver (1995), which, in their
present form, do not reproduce the observed ratio [Zn/Fe]~$\approx0$ 
in Galactic halo stars (see also Hoffman, Woosley, \& Qian 1997).
This immediately raises the question of how accurate the models are in
predicting the relative abundances of Zn and Fe.
Woosley \& Weaver (1995) point out major uncertainties associated with the
position of the mass cut, the time between stalling of the prompt shock 
and neutrino reheating, and the electron distribution in the innermost 
ejected layers.
It has also been suggested that significant amounts of $^{64}$Zn are
synthesized in neutrino-driven winds during the explosions, but the
corresponding yields have so far only been calculated schematically
(Hoffman et al. 1996, 1997). 
Given these uncertainties in the models, it is not clear that there is 
a significant discrepancy between the predicted and observed values of 
[Zn/Fe] in metal-poor stars. 
In any case, we see no reason to treat Zn as an anomalous element and 
no reason to assume that the intrinsic abundance of Zn relative to Fe in 
the DLA galaxies differs from that in the Milky Way. 

To make further progress in discerning the intrinsic nucleosynthesis 
pattern in DLA galaxies, it is critical to obtain observations of  
several weakly depleted elements in the same objects. 
(Conversely, a comparison between weakly and strongly depleted elements 
is required to determine the dust depletion pattern.) In 
principle, N and O are a suitable pair, since they are both weakly 
depleted in the Galactic ISM. The problem in this case  
is that the nucleosynthesis of N is not fully understood. Observations 
of halo stars show considerable scatter 
in [N/O] over the metallicity range -2.5 $\lesssim$ [Fe/H] $\lesssim$ -0.6 
dex, with an average ratio [N/O] $\approx$ -0.2 dex (Laird 1985; Carbon 
et al. 1987). Furthermore, there are uncertainties in the relative 
importance of primary and secondary production of N at low metallicities. 
Significant primary production of N may occur in low-metallicity,  
massive stars (Timmes, Woosley, \& Weaver 1995). These uncertainties cast 
some doubt on the use of [N/O] as a tool for probing the early 
nucleosynthesis history of DLA galaxies. S and Zn are potentially 
more suitable elements, since they are both weakly depleted and 
provide a reliable measure 
of the ratio of $\alpha$-elements to Fe-group elements. In the 
halo stars in our Galaxy with [Fe/H] $\lesssim$ -1, this ratio clearly 
displays the signature of massive stars and associated 
nucleosynthesis in Type II supernovae. We anticipate that further 
observations of N, O, S, and Zn 
will eventually provide a better understanding of nucleosynthesis  
in DLA galaxies.

\acknowledgments

We thank Drs. Nino Panagia, Yichuan Pei, Max Pettini, and Donald York 
for helpful comments on this paper. V. P. K. acknowledges support from 
an AURA Postgraduate Fellowship at STScI.

\clearpage

\figcaption{Logarithmic gas-phase abundances relative to Zn in DLA
galaxies, normalized to solar abundances ({\it top panel}) and 
halo-star abundances ({\it bottom panel}). 
The heavy filled circles and vertical bars show the medians of the 
observed abundances (including upper and lower limits) and the 
standard errors in the medians. The light vertical bars show the full 
range of the actual measurements (excluding upper and lower limits). 
Open circles show the actual measurements, and upward 
pointing triangles show the lower limits for elements measured in 
fewer than three absorbers. The lines show the expected gas-phase 
abundances when the intrinsic (nucleosynthetic) pattern is solar 
({\it top panel}) or halo-star-like ({\it bottom panel}), and when 
the depletion pattern is the same as that in the warm Galactic ISM 
with the indicated values of the dust-to-metals ratio $R/R_{G}$. 
\label{fig1}}

\figcaption{Logarithmic gas-phase abundances relative to Fe in DLA 
galaxies, normalized to solar abundances ({\it top panel}) and 
halo-star abundances ({\it bottom panel}). All symbols have the same 
meaning as in Fig. 1 (with downward pointing triangles indicating 
upper limits). \label{fig2}}

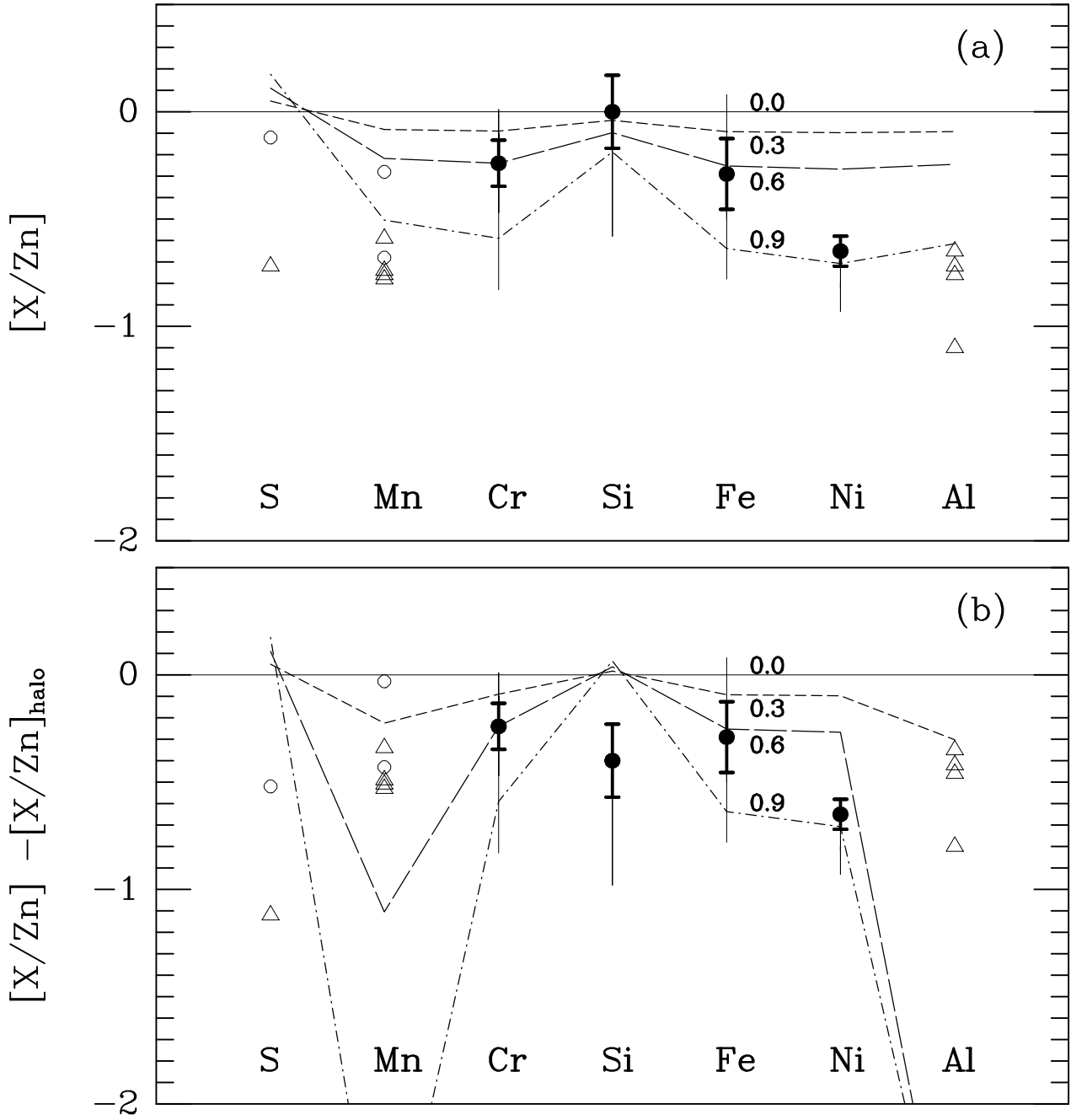
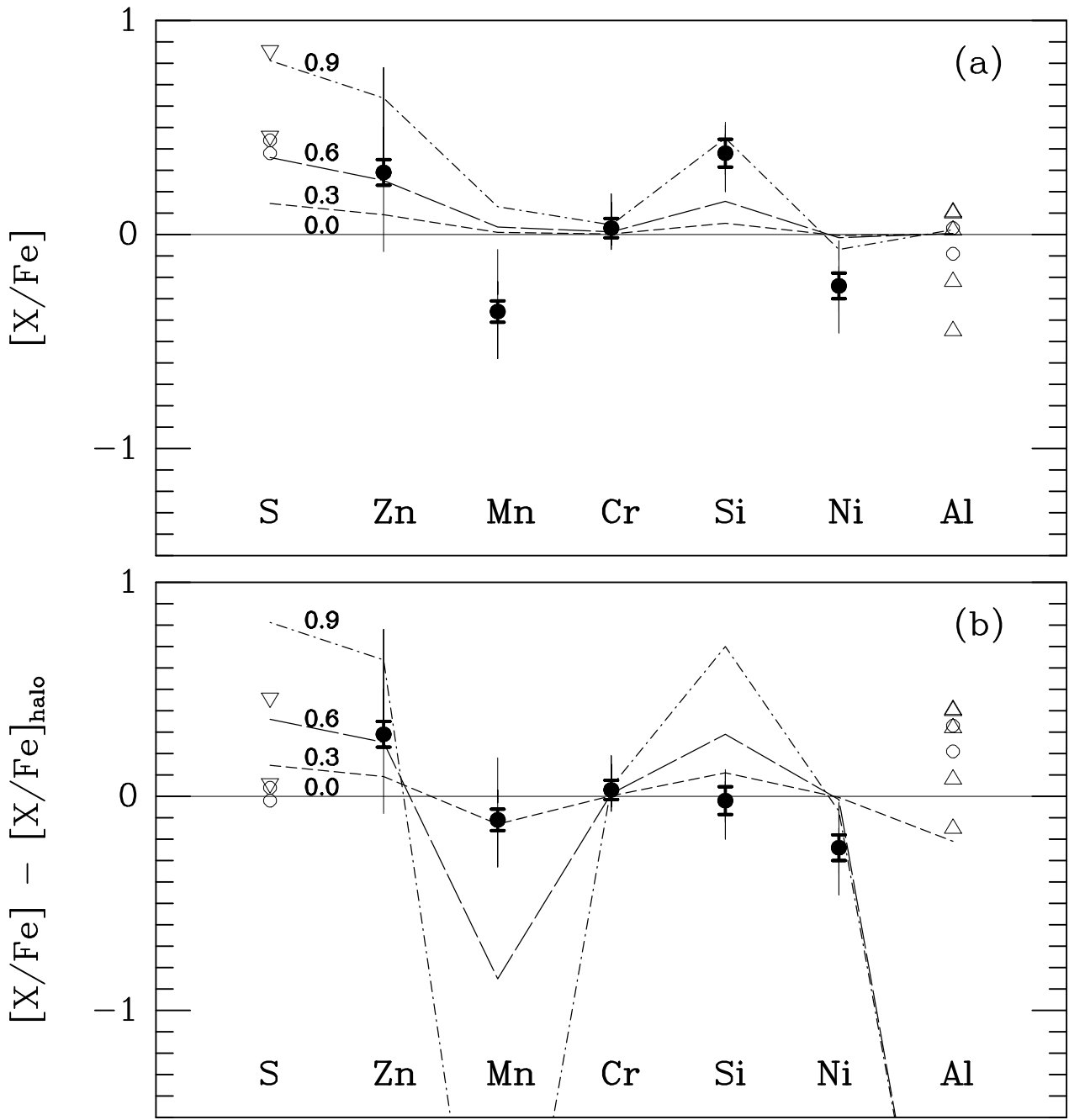
\end{document}